\documentclass[a4paper,11pt]{article}
\pdfoutput=1 

\usepackage{jinstpub} 

\title{\boldmath The Detectors of the Mu2e Experiment}

\author{S. Giovannella}
\author[1]{on behalf of the Mu2e collaboration%
  \note{https://mu2e.fnal.gov/mu2e\_collaboration\_list.shtml}}
\affiliation{Laboratori Nazionali di Frascati dell'INFN\\
  Via Enrico Fermi 40, 00044, Frascati (Roma), Italy}

\emailAdd{simona.giovannella@lnf.infn.it}

\abstract{
  The Mu2e experiment aims to test Charge Lepton Flavour Violation
  to an unprecedented level, enhancing the current sensitivity by four
  orders of magnitude for the neutrinoless conversion of muons into
  electrons. A series of graded solenoids convey an intense, pulsed
  muon beam to an aluminum target. The main detector components are
  a low mass straw drift tubes tracker, a pure Cesium Iodide calorimeter
  and an extruded plastic scintillator cosmic ray veto. Requirements,
  tests on prototypes and status of the production will be discussed.
}


\proceeding{15$^{\text{th}}$ Topical Seminar on Innovative Particle and Radiation Detectors\\
  14-17 October 2019\\
  Siena, Italy}

\begin{document}
\maketitle
\flushbottom

\section{The Mu2e Experiment}
\label{sec:intro}

The Mu2e experiment \cite{TDR} at Fermilab will search for the
charged-lepton flavor violating neutrino-less coherent conversion
of a negatively charged muon into an electron in the field of an
aluminum nucleus. The process produces a mono-energetic electron
with an energy slightly below the muon rest mass (104.967 MeV). If
no events are observed, Mu2e will set a
limit on the ratio between the conversion rate and the muon capture
rate of $R_{\mu e}$~$\leq 8.4\ \times\ 10^{-17}$ (@ 90$\%$ C.L.).
This will improve the current limit \cite{Sindrum-II} by four
orders of magnitude.
On the other hand, an observation of Charge Lepton Flavour Violation
(CLFV) events will provide a clear indication of New Physics (NP)
beyond the Standard Model up to mass scales of nearly $10^4$ TeV,
far beyond the direct search reach at colliders, complementing and
extending other CLFV searches on a wide range of NP scenarios
\cite{CLFV-theory}.

\begin{figure}[ht]
  \begin{center}
    \includegraphics[width=1.0\textwidth]{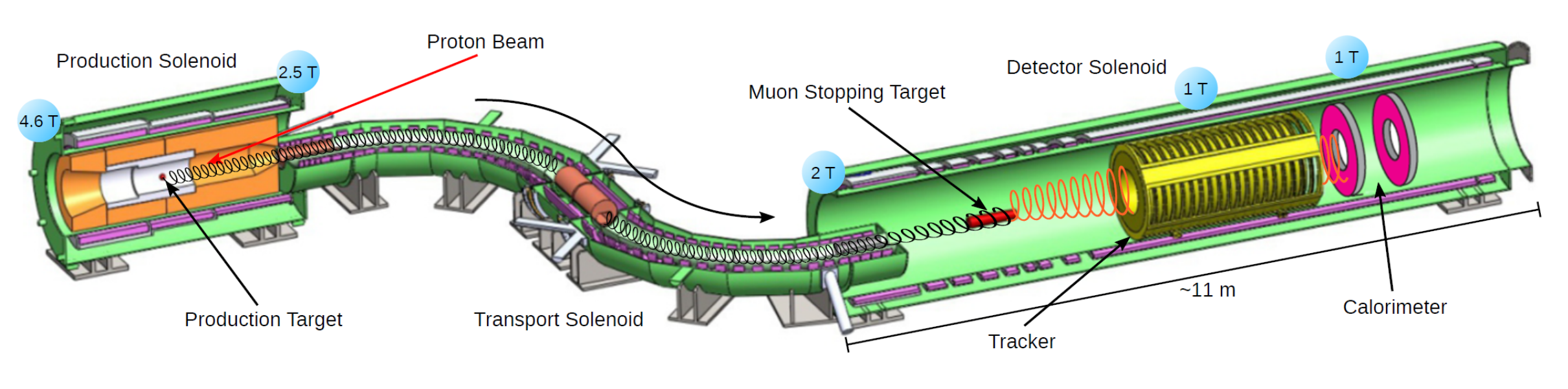}
  \end{center}
  \caption{The Mu2e experiment. Cosmic Ray Veto and Stopping
    Target Monitor are not shown.}
  \label{Fig:Mu2e}
\end{figure}

The Mu2e design is based on the MELC concept \cite{MELC}. An
intense pulsed muon beam ($\sim 10^{10} \mu/$sec) is produced by
8 GeV, 8 kW protons hitting a tungsten target and it is stopped
on an aluminum target after travelling inside a very long, curved
series of solenoids (Fig.~\ref{Fig:Mu2e}).
The strong negative gradient of the Production Solenoid, from 4 to
2.5 T, confines soft pions and increases the yield through magnetic
reflection. The S-shaped Transport Solenoid efficiently transfers
low energy, negatively charged particles while allowing a large
fraction of pions to decay into muons.
The Detector Solenoid has a graded field from 2 to 1 Tesla in
the upstream region of the stopping target to increase acceptance
for Conversion Electron (CE) events. 

The Mu2e detector, just downstream of the aluminum target inside
a 1T solenoid, is composed of a tracker and an electromagnetic
calorimeter. The Mu2e tracker measures the momentum of the conversion
electron and separates it from the background. The crystal
calorimeter plays an important role in providing particle
identification capabilities and a fast online trigger filter, while
also aiding the track reconstruction capabilities. The detector
solenoid is in vacuum, at $10^{-4}$ Torr, and in a high radiation
environment. The entire detector region and part of the transport
solenoid are surrounded by a Cosmic Ray Veto (CRV) that reduces the
cosmic ray background.
A High Purity Germanium Detector and a Lanthanum Bromide crystal
constitute the  Stopping Target Monitor, placed $\sim 35$ m after
the stopping target, which provides normalization to CLFV events by
detecting $\gamma$-rays emitted from muon capture in the aluminum
target.

In order to reach the required sensitivity, control of the background
to the level of less than 0.5 expected events is required.
The background coming from the
beam is reduced by means of a pulsed beam structure with a proton
extinction lower than 10$^{-10}$: a delay in the start of the live
window of $\sim$ 700 ns after the bunch arrival time removes the
prompt background from the acquired data.
The extinction level is monitored by detecting scattered protons from
the production target to evaluate the fraction of out-of-time beam.

\section{The tracking system}
\label{sec:tracker}

\begin{figure}[!th]
  \centering
  \begin{tabular}{cc}
    \includegraphics[width=0.5\textwidth]{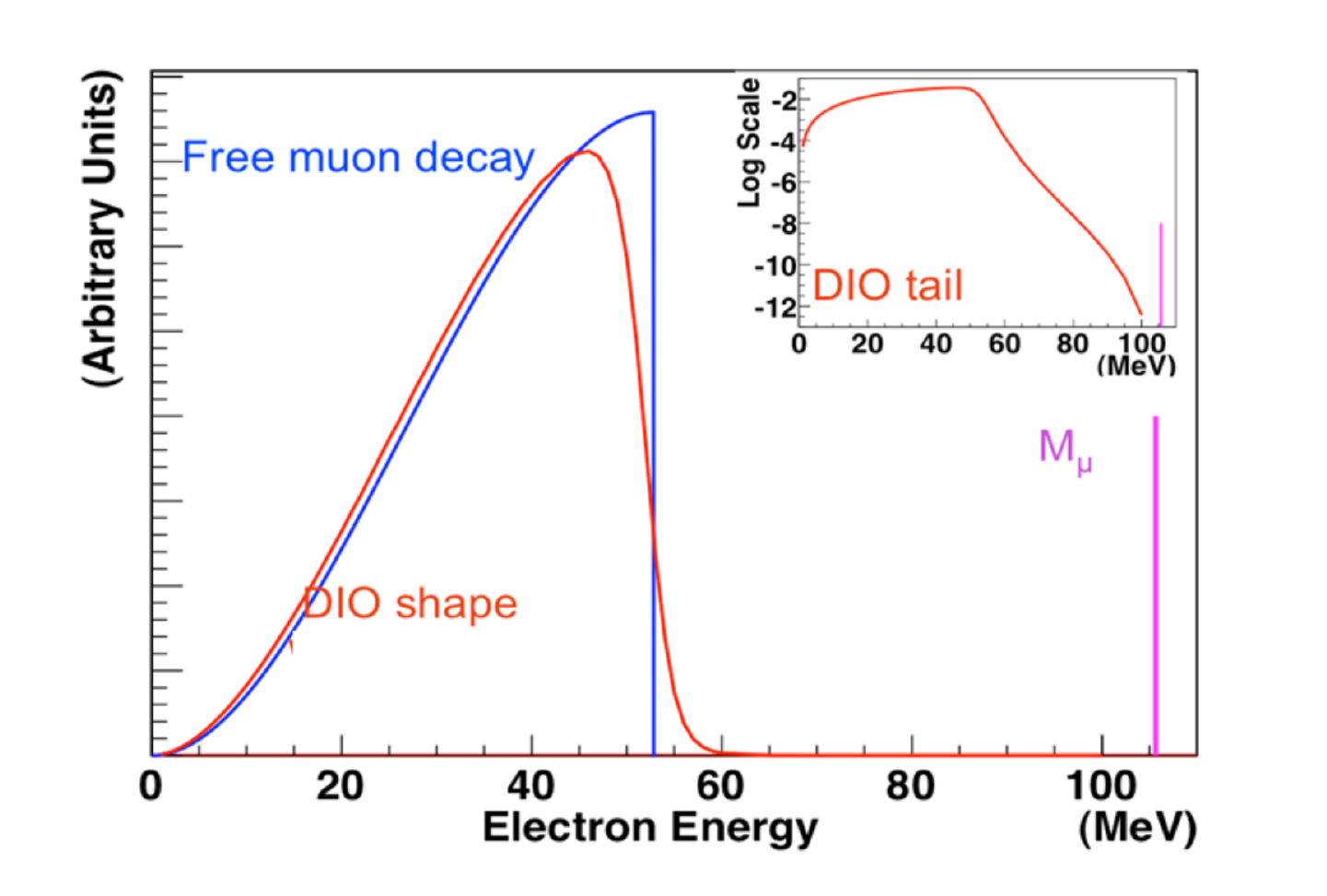} &
    \includegraphics[width=0.5\textwidth]{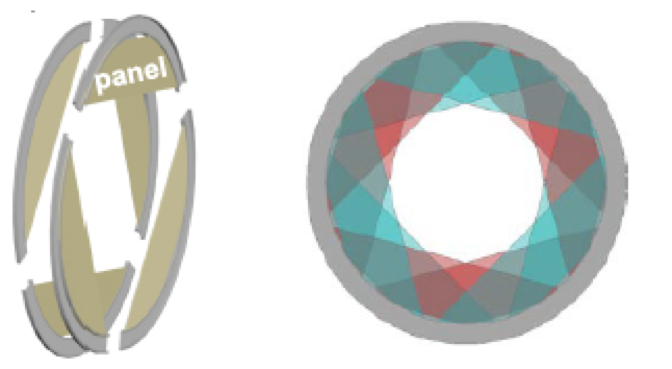} \\
  \end{tabular}
  \caption{Left: energy spectrum for electrons produced from
    free muon decays (blue), muon decays in orbit (red) and
    conversion electrons (purple). Right: Sketch of the Mu2e
    straw tracker system. The basic element is the panel, where
    straws are organized in two staggered layers. Six panels
    arranged as shown in in the middle figure above form a plane;
    two planes rotated by $30^\circ$ constitute a station, right.
    The tracker, containing 18 stations, is 3.2 meters long.}
  \label{Fig:tracker}
\end{figure}

The Mu2e tracker system \cite{Tracker} is designed to maximize
acceptance for conversion electrons while minimizing the contamination
from the muon Decay-In-Orbit (DIO) background, where nuclear
modifications push the DIO spectrum towards the CE signal
(Fig.~\ref{Fig:tracker} left). Energy loss and detector resolution
produce an overlap of the two processes. The selected design is based
on nearly 20,000 low mass straw drift tubes of 5 mm in diameter,
with a 15 $\mu$m Mylar wall and 25 $\mu$m sense wire.
Straws of lengths ranging from 430 to 1220 mm are oriented transversely
to the solenoid axis and arranged in 18 stations (Fig.~\ref{Fig:tracker}
right), for a total length of 3.2 metres along the solenoid axis. A
central hole, 38 cm in diameter, makes the device blind to low momentum
background particles ($p<55$ MeV/c) which are constrained to low radius
by the solenoidal field.

\begin{figure}[!th]
  \vspace{0.3cm}
  \centering
  \includegraphics[width=1.0\textwidth]{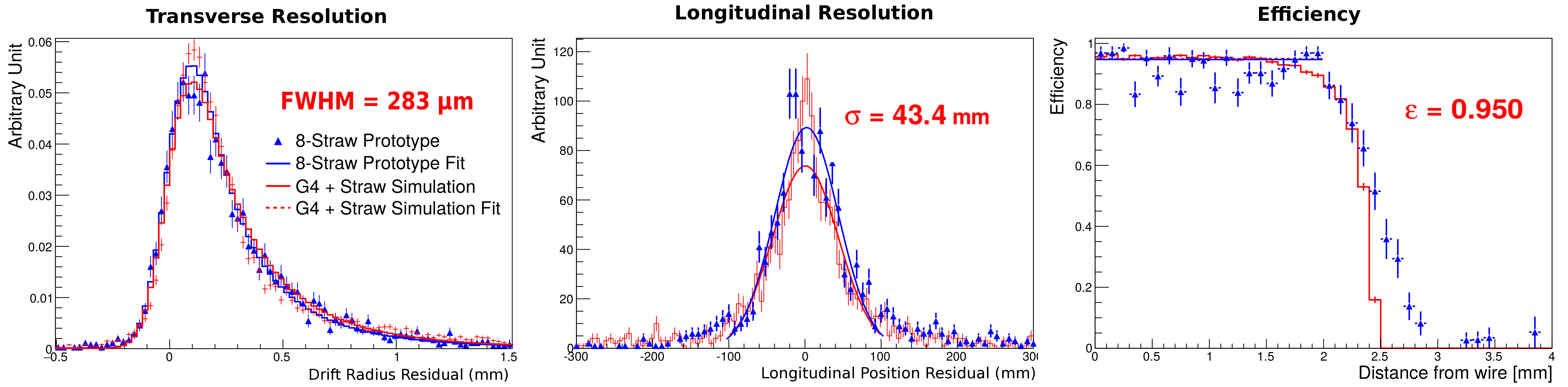}
  \caption{Longitudinal (left), transverse (center) position
    resolution and efficiency (right) for an eight channel
    prototype of the tracker.
    Data from minimum ionizing particles (blue triangles) are
    compared with Monte Carlo simulation (red crosses). 
    Resolution is extracted with Gaussian fits to the spectra.}
  \label{Fig:trk-proto}
\end{figure}

An eight channel tracker prototype was built and tested with
cosmics rays to measure performances and tune detector simulations.
In Fig.~\ref{Fig:trk-proto}, the position resolution and straw
efficiency are compared with Monte Carlo expectations.
Good reproducibility of data is observed.

\begin{figure}[!th]
  \vspace{0.3cm}
  \centering
  \includegraphics[width=0.5\textwidth,height=6cm]{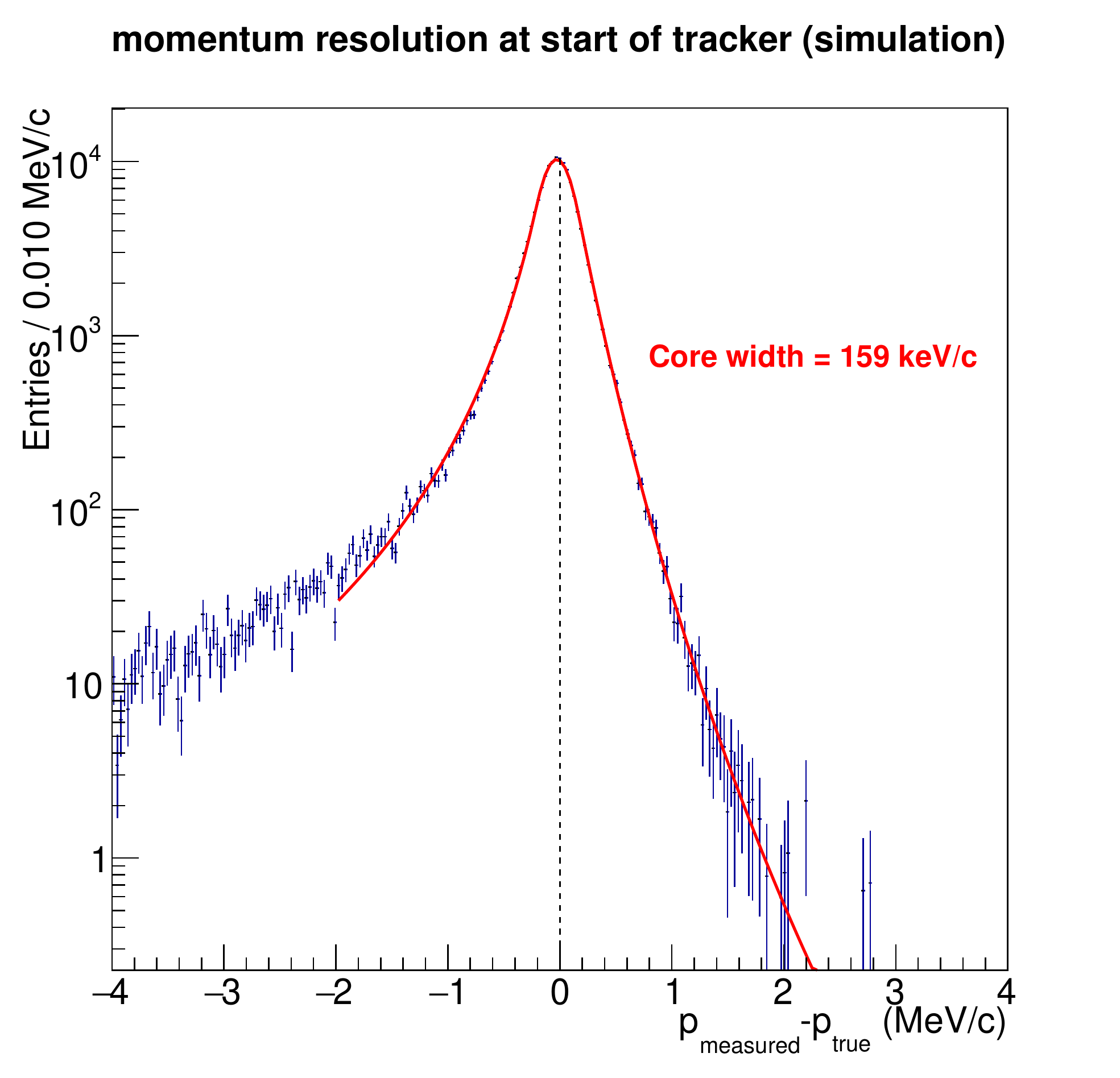}
  \caption{Momentum resolution evaluated with the fully tuned
    Mu2e simulation.}
  \label{Fig:trk-reso}
\end{figure}

The tracker performance is studied with Monte Carlo using the
full Mu2e simulation. Results are reported in Fig.~\ref{Fig:trk-reso}.
The core momentum resolution of 159 keV/c is well within physics
requirements and stable when increasing accidental hit rate.
The total track efficiency of $\sim 9\%$ is fully dominated by
geometric acceptance.

\begin{figure}[!th]
  \centering
  \includegraphics[width=\textwidth]{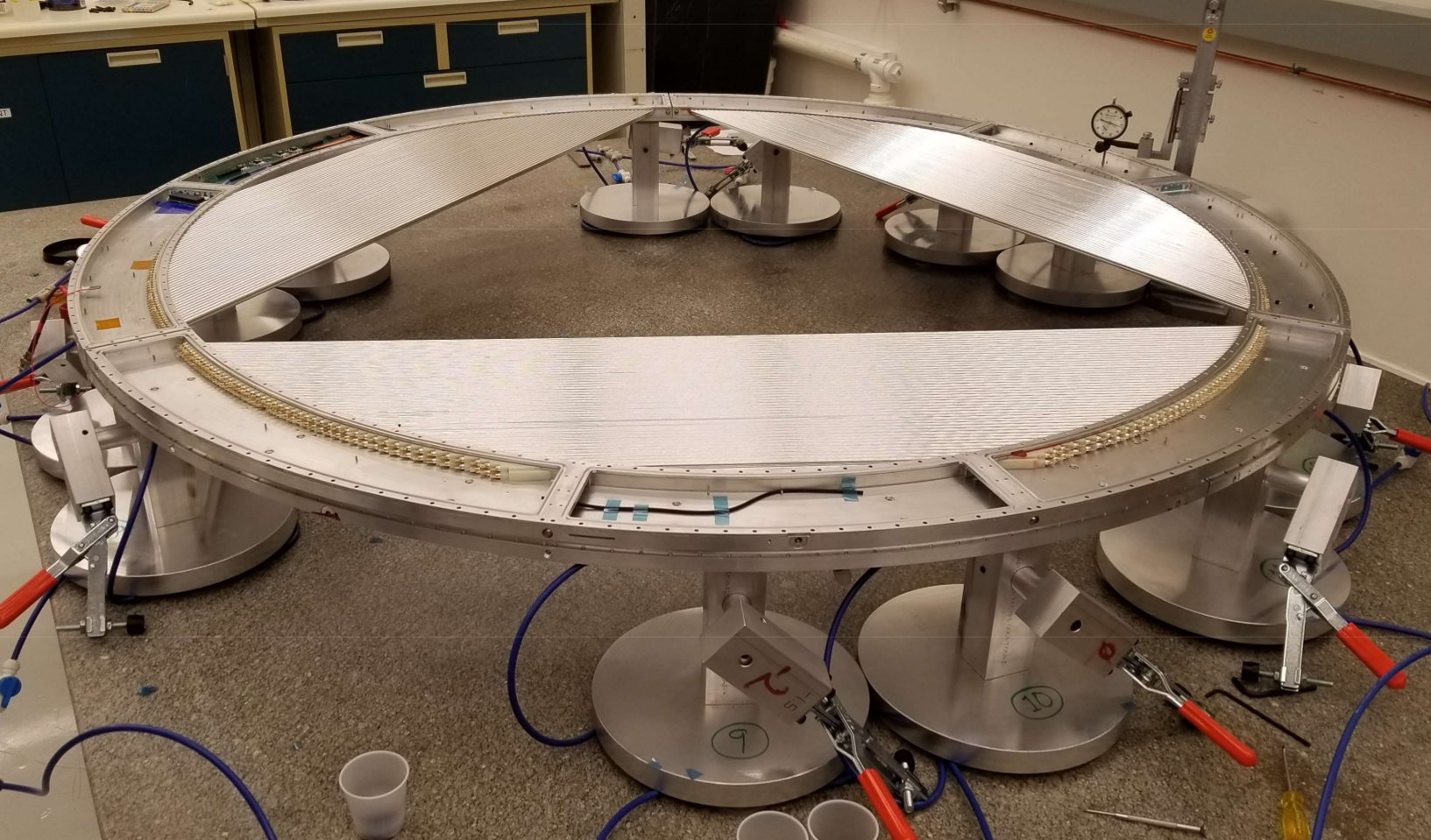}
  \caption{Tracking plane being assembled with pre-production
    panels.}
  \label{Fig:trk-plane}
\end{figure}

At the moment of writing, twelve pre-production panels are under
construction and testing. In Fig.~\ref{Fig:trk-plane}, three
panels are assembled to form a tracking plane.
A vertical slice test on fully instrumented panels with the entire
Front-End Electronics chain will be performed.

\section{The calorimeter system}
\label{sec:calo}

\begin{figure}[!th]
  \begin{center}
      \includegraphics[width=1.0\textwidth]{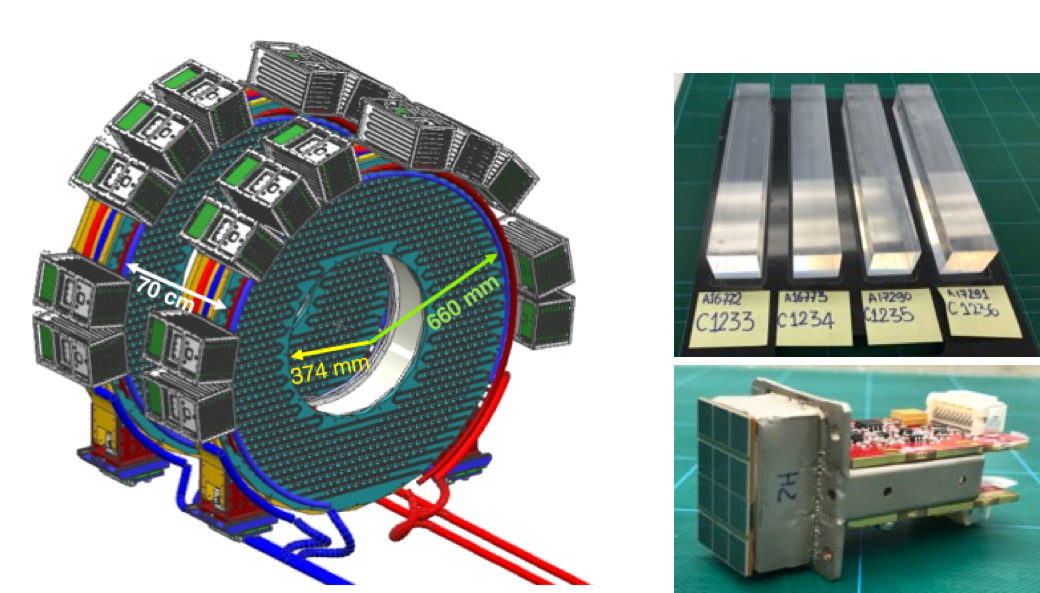}
  \end{center}
  \caption{Left: sketch of the calorimeter system. The cooling
    pipes and the on board racks containing the DIRAC boards
    are visible. Right: pure CsI calorimeter crystals (top) and
    a readout channel, composed by two UV-extended SiPMs and the
    corresponding analog FEE boards (bottom).}
\label{Fig:calo} 
\end{figure}

The Mu2e calorimeter \cite{TDR-EMC} has to provide confirmation for
CE signal events, a powerful $e/\mu$ separation - with a muon
rejection factor of $\sim 200$, a standalone trigger and seeding for
track reconstruction. An energy resolution of $O(10\%$) and a time
resolution of $500$ ps for 100 MeV electrons are sufficient to fulfil
these requirements.
The calorimeter design consists of two disks made from 674 undoped
CsI scintillating crystals with ($34\times 34\times 200$) mm$^3$
dimension. Each crystal is read-out by two custom array large area
($2\times 3$ of $6\times 6$ mm$^2$ cells) UV-extended Silicon
Photo-Multipliers (SiPMs).  Each SiPM is connected to a Front-End
Electronics (FEE) board providing amplification and shaping of the
signal. Groups of 20 signals are sent to a custom digitizer module
(DIRAC, DIgitizer and ReAdout Controller) where they are sampled at
200 Mega samples per second and transferred to the Mu2e data acquisition
system. A radioactive source and a laser system allow setting the energy
scale and monitor the fast changes of response and resolution.
The crystals will receive an ionizing dose of 90 krad and a fluence
of $3\times10^{12}$ n/cm$^2$.
The photosensors,
being shielded by the crystals, will get a three times smaller dose.
The layout of the calorimeter system and pictures of crystals and a
readout channel are shown in Fig.~\ref{Fig:calo}.

\begin{figure}[!th]
  \vspace{0.3cm}
  \centering
  \begin{tabular}{cc}
  \includegraphics[width=0.5\textwidth]{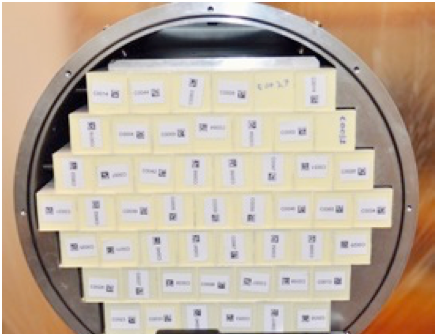} &
  \includegraphics[width=0.5\textwidth]{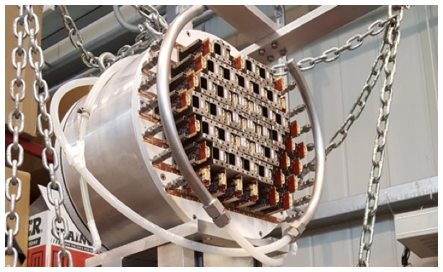} \\
  \end{tabular}
  \caption{Module-0, a large size prototype of the Mu2e calorimeter.
    Left: front view before mounting the source panel, where the
    staggered crystal structure is visible. Right: rear side, with
    readout channels and cooling circuit.}
  \label{Fig:emc-mod0}
\end{figure}

\begin{figure}[!t]
  \centering
  \begin{tabular}{ccc}
  \includegraphics[width=0.33\textwidth]{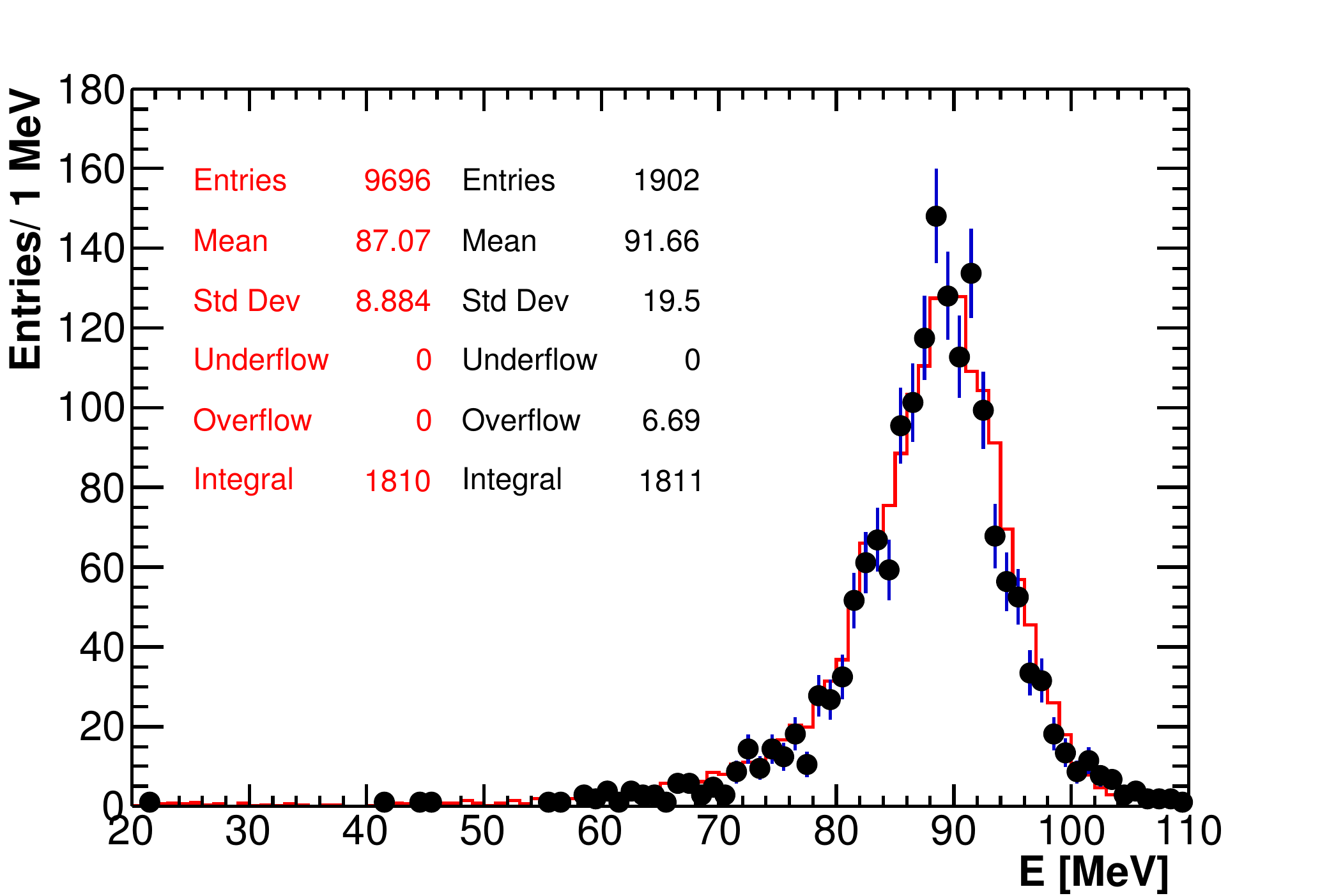}  &
  \includegraphics[width=0.33\textwidth]{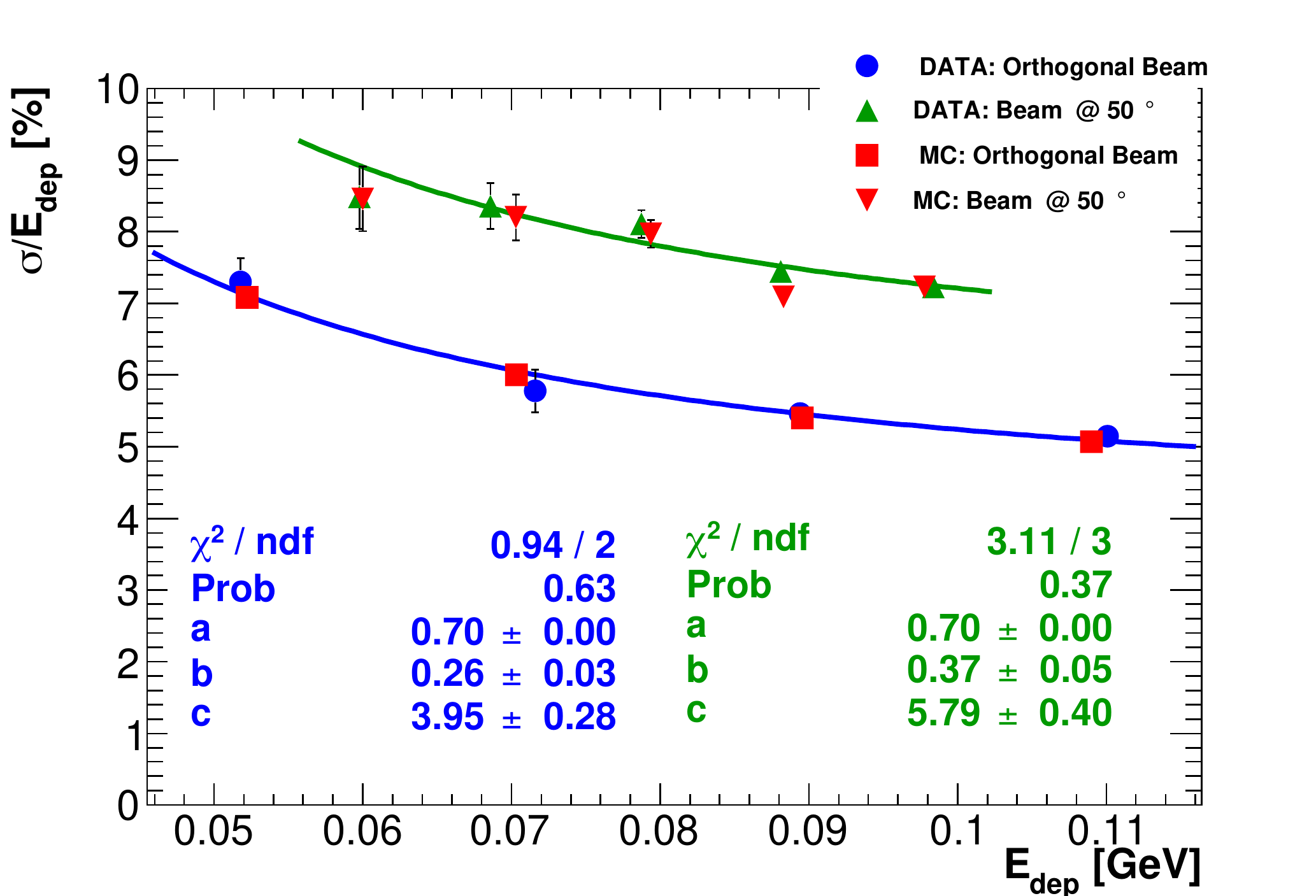} &
  \includegraphics[width=0.33\textwidth]{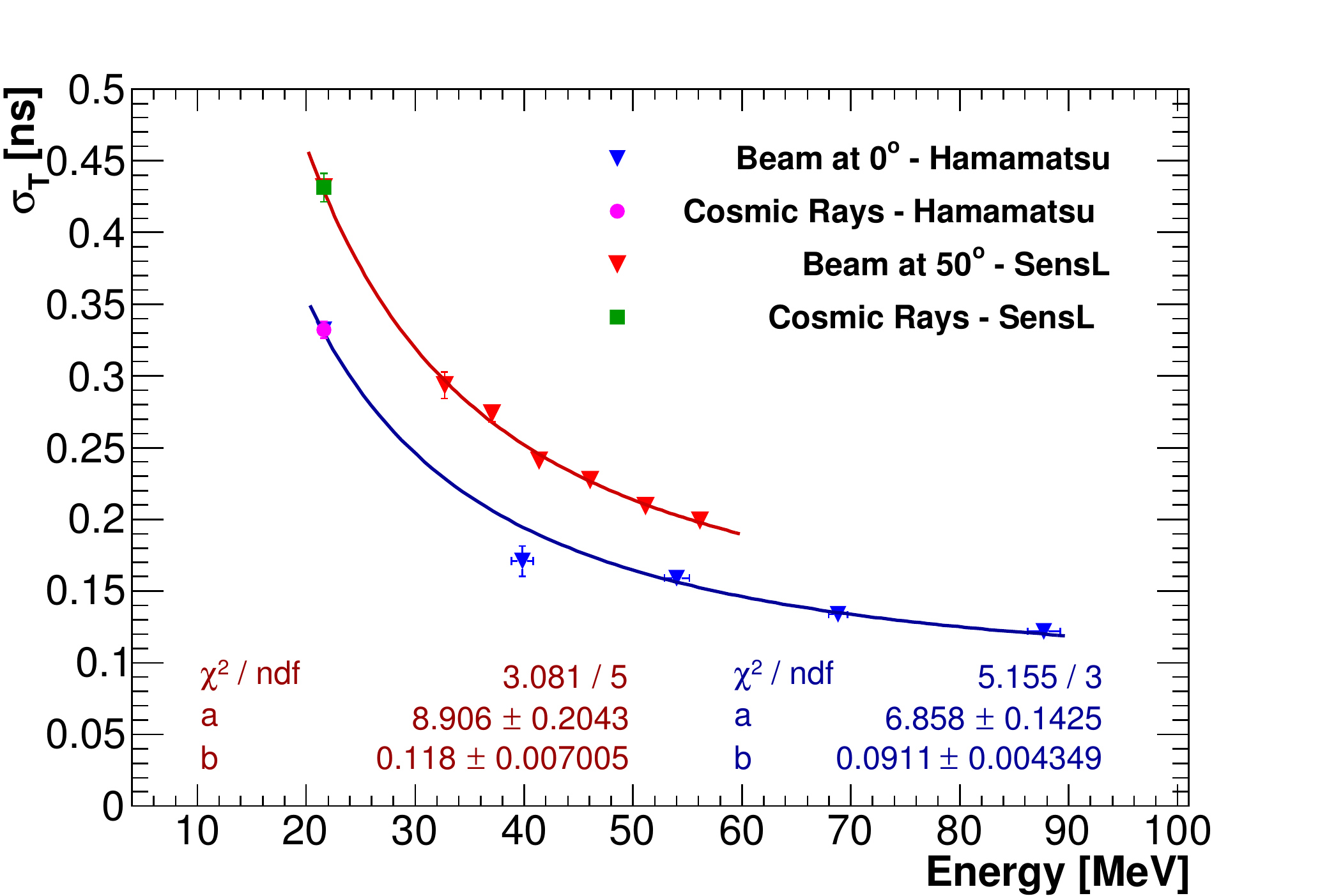} \\
  \end{tabular}
  \caption{Calorimeter performance evaluated with a large-scale prototype,
    Module-0, using 60-120 MeV electron beam. Left: data-MC comparison of
    the energy distribution for 100 MeV beam. Energy (center) and time
    (right) resolution for orthogonal and 50$^\circ$ impinging electrons.
    The energy resolution is compared with results expected from simulated
    data.}
  \label{Fig:calo-testbeam}
\end{figure}

A long R\&D phase with small prototypes demonstrates that the calorimeter
design easily satisfies the requirements
\cite{NIM-LYSO1,NIM-LYSO2,NIM-BaF2,Proto-EMC}.
Pre-production components have been used to build a large size calorimeter 
prototype, Module-0 (Fig.~\ref{Fig:emc-mod0}), with 51 crystals and 102
SiPMs and front end boards \cite{Module-0}. It represents a portion of the
final disk and has been used to test the integration and assembly procedures
and to evaluate the operations of running in vacuum and at low temperatures.
Module-0 performance was tested with an electron beam of 60-120 MeV at the
INFN Beam Test Facility in Frascati \cite{BTF}. The energy distribution for
100 MeV electrons is well reproduced by the calorimeter simulation,
Fig.~\ref{Fig:calo-testbeam} left. Energy and time resolution are evaluated
with particles impinging on the calorimeter surface both at 0 and 50 degrees.
The latter is the expected incidence angle for conversion electrons in Mu2e.
An energy resolution of 5\% (7\%) and a time resolution of 120 ps (150 ps)
are obtained for 100 MeV particles impinging at $0^{\circ}$ ($50^{\circ}$),
Fig.~\ref{Fig:calo-testbeam} center and right. Results satisfy
physics requirements and are well reproduced by simulation.

\begin{figure}[!t]
  \centering
  \includegraphics[width=\textwidth]{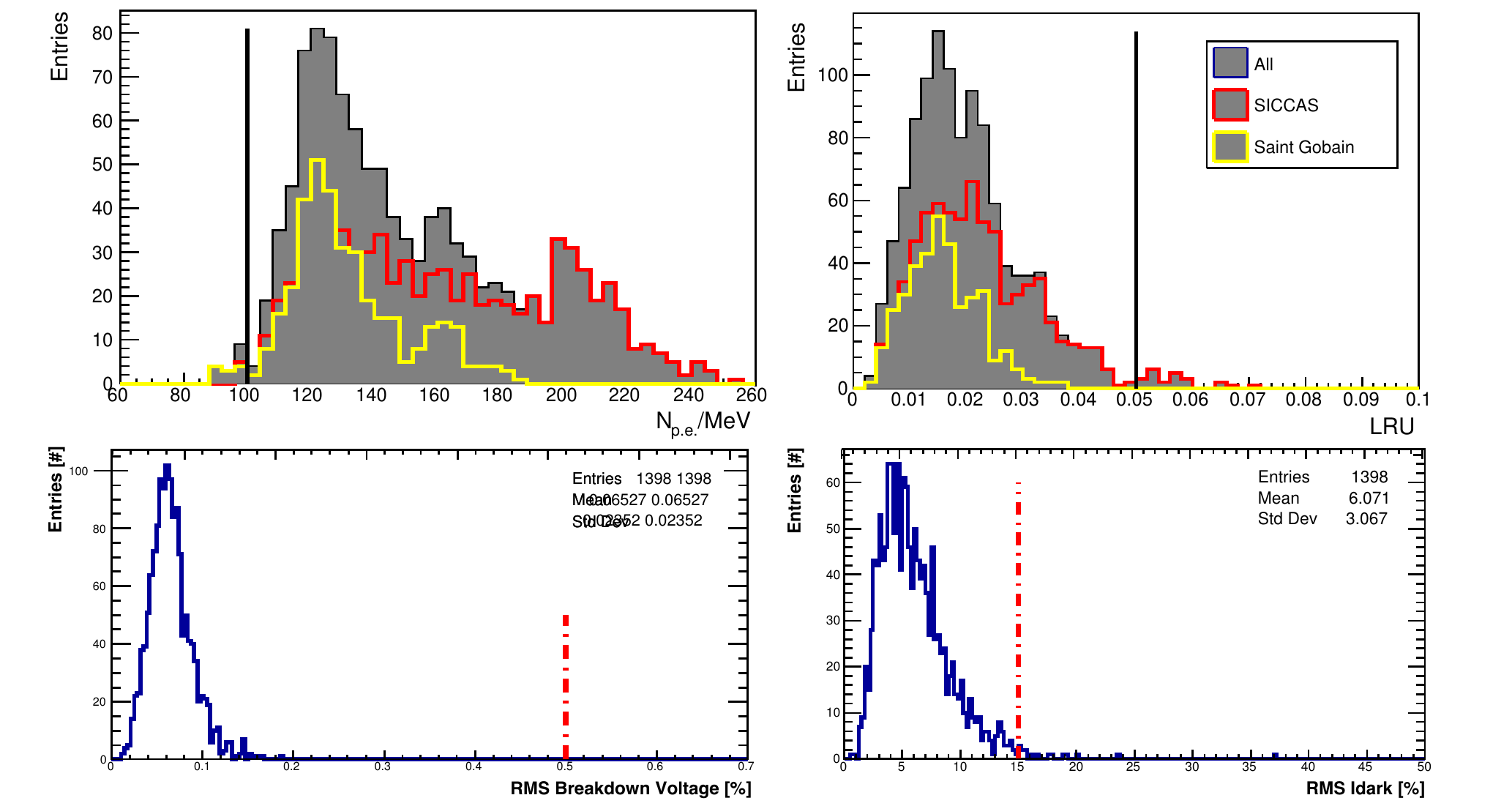}
  \caption{Summary of Quality Control measurements for production
    CsI crystals and Silicon Photo-Multipliers: CsI light yield
    (top left) and longitudinal response uniformity (top right);
    RMS of the breakdown voltage (bottom left) and of the dark
    current over the six cells of SiPMs (bottom right).
    Vertical lines represent the Quality Control acceptance cuts.}
  \label{Fig:calo-qc}
\end{figure}

The complete production components for SiPMs and 85\% of production
crystals have been received and characterized. For all of the 4000
sensors, the breakdown voltage and the dark current are measured
at different temperatures. The spread of these quantities over
the six cells of each sensor is used as quality control parameter
(Fig.~\ref{Fig:calo-qc} bottom). The overall rejection factor is
1.2\%, dominated by those sensors whose dark current RMS is too large.
The Quality Control of CsI crystals foresees a dimensional control,
with 0.1 mm tolerance with respect to nominal values, and a
measurement of the optical properties \cite{QAcrystals}.
In Fig.~\ref{Fig:calo-qc} (top) the number of photoelectrons and
the uniformity response along the crystals are reported for both
of the CsI producers. About 10\% of the crystals have been rejected,
mostly due to problems with mechanical tolerances.
Irradiation tests have been carried out for small CsI and SiPM
production subsamples. Results show that the calorimeter will be
able to operate at the end of the Mu2e lifetime at a temperature
below $0^\circ$ C.
Mean Time To Failure tests on photosensors demonstrate an MTTF
value 10 times larger than the experiment needs.

The prototypes of FEE and DIRAC have been exposed to a large ionization
dose and neutron fluence to qualify rad-hard components. A slice test
with the whole calorimeter electronic chain provides results comparable
to those achieved using a commercial digitizer. A DIRAC prototype is
currently used to read 16 channels of Module-0.

\section{Cosmic Ray Veto}
\label{sec:crv}

\begin{figure}[!th]
  \vspace{0.5cm}
  \centering
  \begin{tabular}{cc}
  \includegraphics[width=0.5\textwidth]{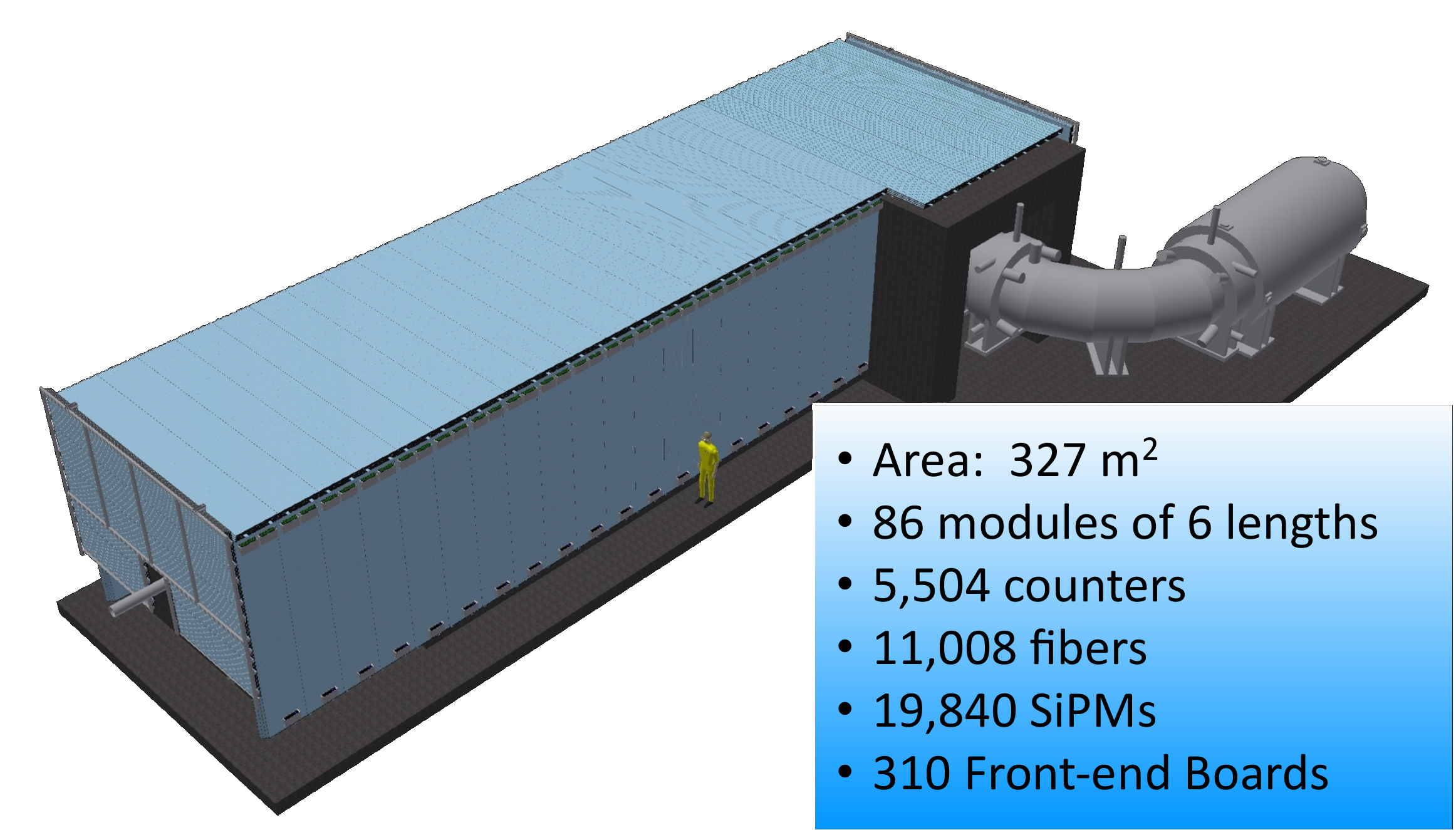} &
  \includegraphics[width=0.5\textwidth]{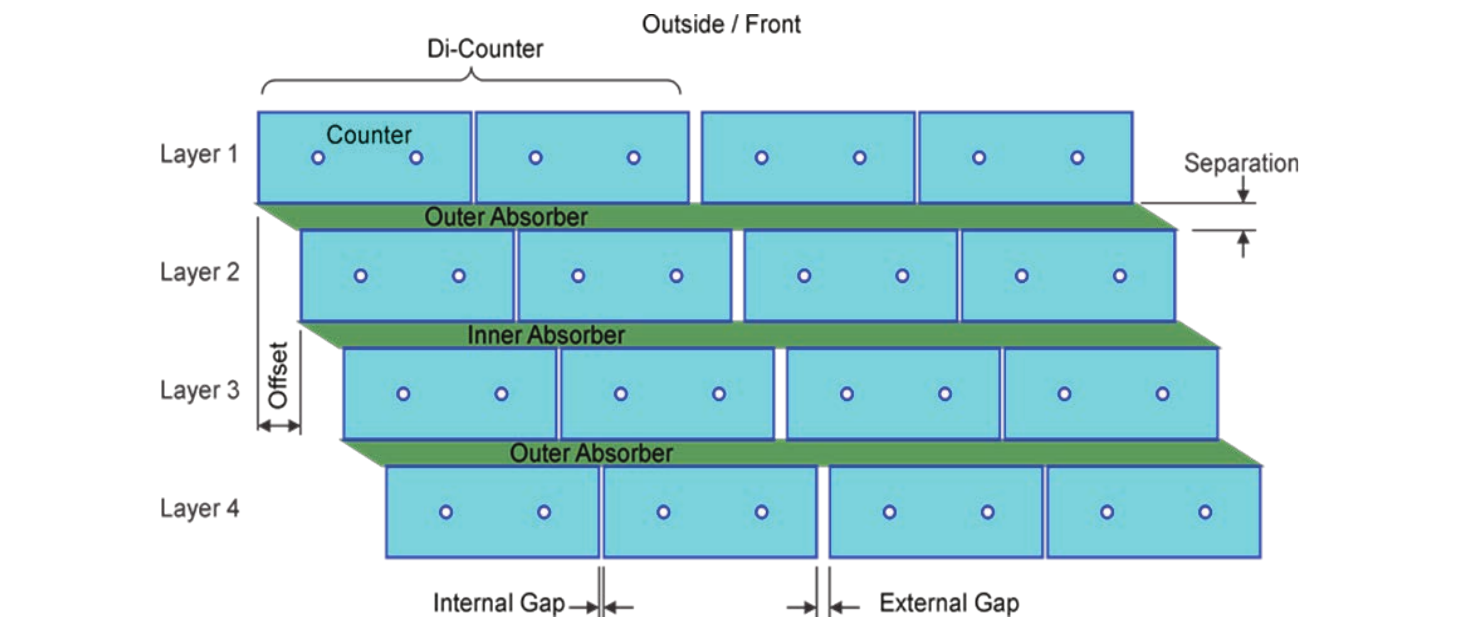} \\
  \end{tabular}
  \caption{Left: the Mu2e Cosmic Ray Veto system, covering
    the detector solenoid and part of the transport solenoid.
    Right: layout of a CRV module.}
  \label{Fig:crv-design}
\end{figure}

In absence of the vetoing system, cosmic ray muons interacting with
the detector materials produce false signal CE candidates at a rate
of approximately one/day. In order to maintain the background under
the required level, the CRV has to provide a vetoing efficiency of at
least 99.99\% for cosmic ray tracks while withstanding an intense
radiation environment.
The Cosmic Ray Veto system \cite{CRV} is made by four staggered layers
of extruded plastic scintillation counters with two embedded 1.4 mm
diameter Wavelength Shifting Fibers/counter, alternated with absorber
slabs (Fig.~\ref{Fig:crv-design}).
Each fiber is readout by means of 2$\times$2 mm$^2$ SiPMs. To achieve
the required coverage, a total of 5,504 counters are needed, organized
in 86 modules of six different lengths for a total surface coverage of
327 m$^2$.

\begin{figure}[!t]
  \centering
  \includegraphics[width=\textwidth]{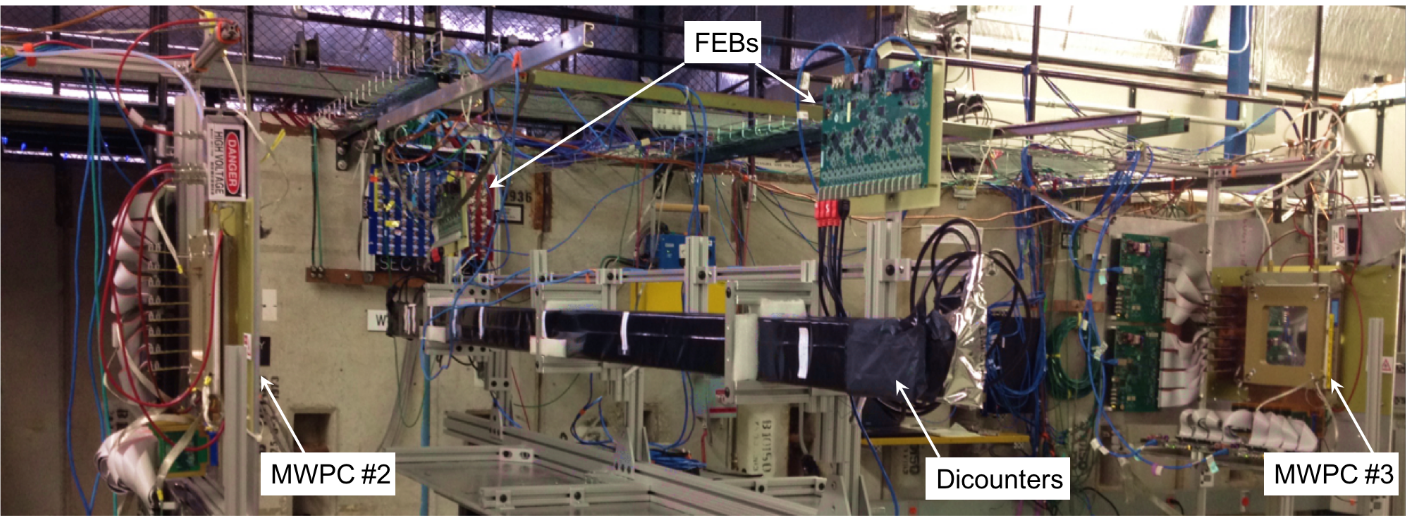}
  \caption{Set-up of the CRV test beam. Protons are tracked with
    multi-wire proportional chambers. Front-End Boards (FEB) are
    visible on the top of the counter.}
  \label{Fig:crv-testbeam}
\end{figure}

\begin{figure}[!t]
  \vspace{0.5cm}
  \centering
  \begin{tabular}{cc}
  \includegraphics[width=0.5\textwidth]{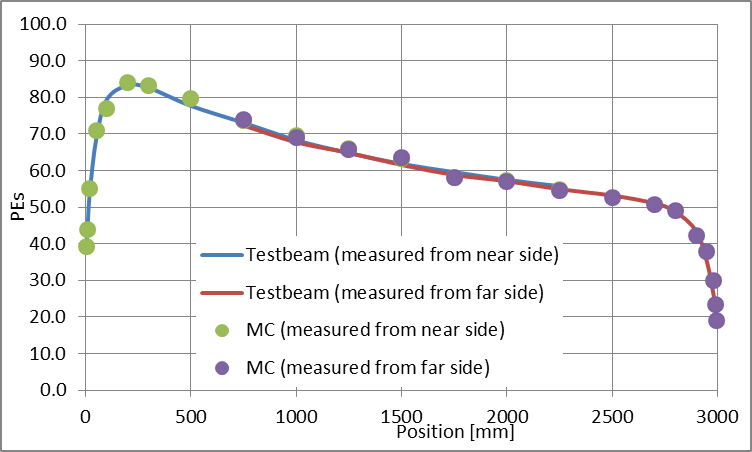} &
  \includegraphics[width=0.5\textwidth]{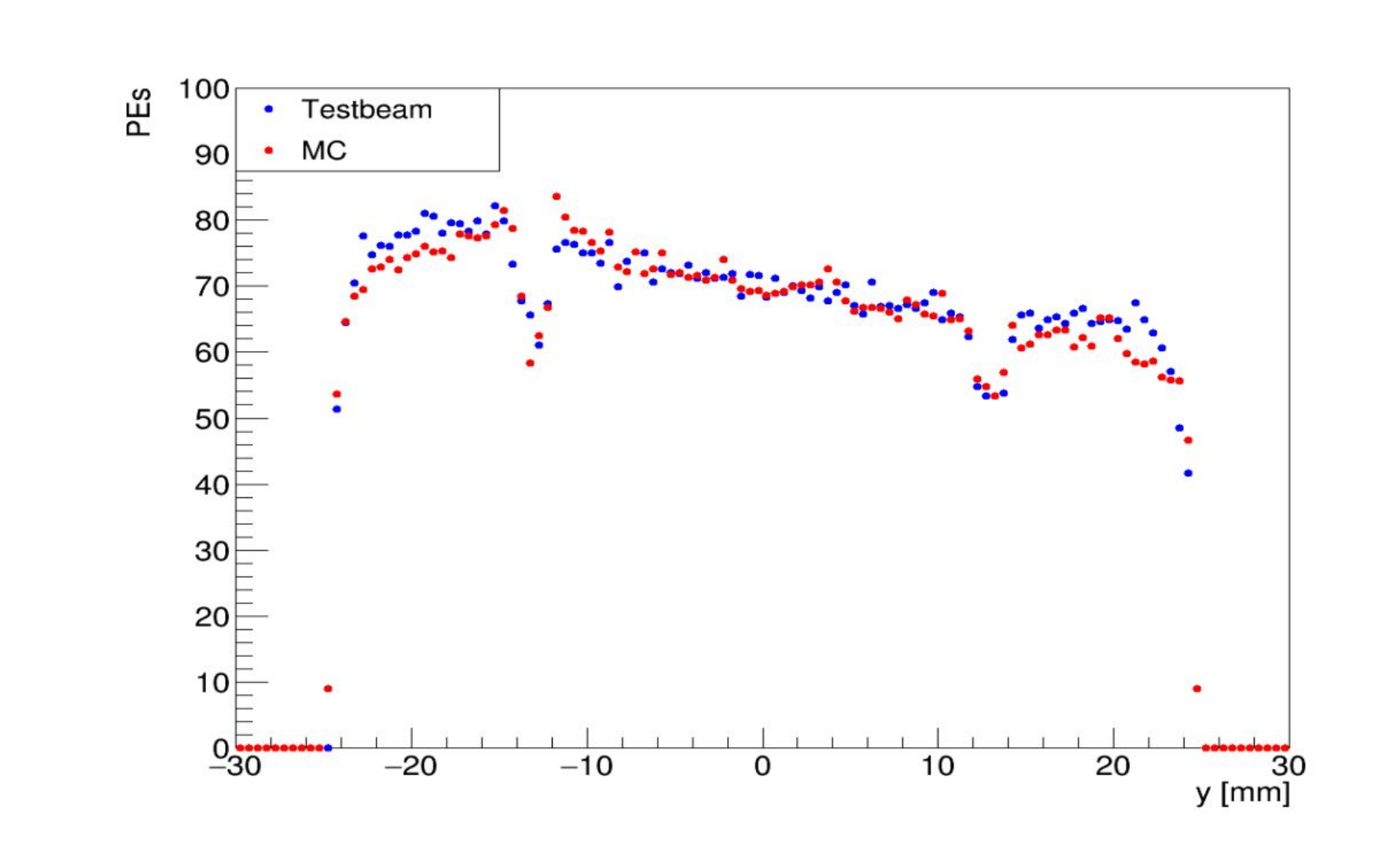} \\
  \end{tabular}
  \caption{Comparison of number of photo-electrons between simulated
    and test beam data for 120 GeV protons normally incident at
    different locations along (left) and across (right) a CRV
    counter.}
  \label{Fig:crv-pe}
\end{figure}

Measurements on a full size prototype with 120 GeV protons in the
Fermilab test beam area was carried out (Fig.~\ref{Fig:crv-testbeam})
demonstrating that the needed light yield can be reached: the number
of photo-eletrons obtained at 1 meter from the readout end provides
a safety factor of $\sim 40\%$ with respect to the requirements
\cite{CRVnpe}.
In Fig.~\ref{Fig:crv-pe} test beam results are compared with the
results obtained from the CRV counter simulation, which includes
scintillation and Cerenkov photon production/transport, SiPM and
electronics responses. Good agreement is obtained after tuning
the Monte Carlo parameters.
Irradiation of CRV SiPMs with neutrons was also tested to
understand the maximum level of fluence acceptable for operations
\cite{CRVradhard}: neutrons could deteriorate the sensors response
and increase the detector occupancy and dead-time so that shielding
is mandatory.

\begin{figure}[!thb]
  \begin{center}
    \includegraphics[width=\textwidth]{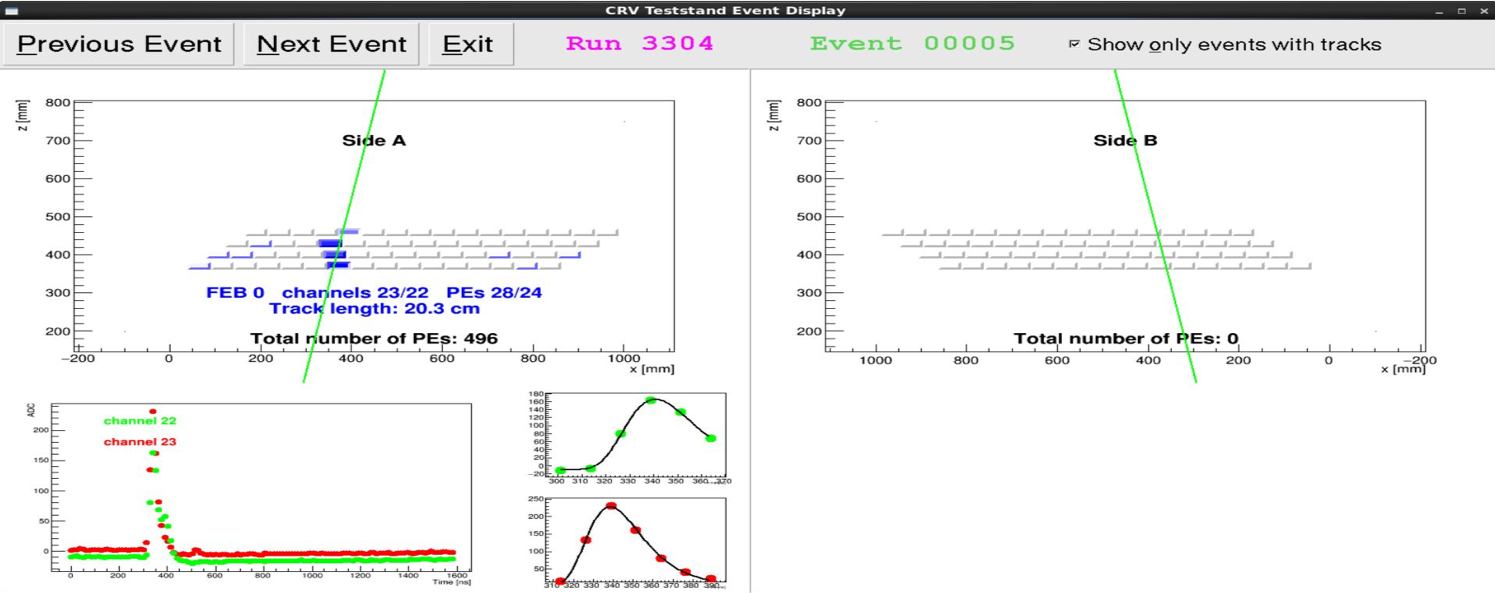} 
  \end{center}
  \caption{Example of event display at the cosmic ray test stand
    used to qualify CRV modules.} 
  \label{Fig:crv-display} 
  \vspace{1cm}
\end{figure}

The assembly of CRV di-counters started in June 2018 and about
half of them have been produced. Production of photosensors and
electronics are also underway and 6\% of the modules have been
assembled.
A test stand with cosmic rays is used to control the modules after
production. An example of a cosmic ray event, as recorded by the
test stand and by the CRV module under test, is shown in
Fig.~\ref{Fig:crv-display}.


\section{Conclusions and perspectives}
\label{sec:theend}

The Mu2e experiment will exploit the world's highest intensity muon
beams of the Fermilab Muon Campus to search for CLFV, improving current
sensitivity by a factor $10^4$ and with a discovery capability over a
wide range of New Physics models.
A low mass straw tube tracker, a pure CsI crystal calorimeter with
SiPM readout and a high efficiency cosmic ray veto have been selected
to satisfy the demanding requirements.
Tests on prototypes and pre-production modules meet the experimental
needs.
Detector construction is in progress and is expected to be completed
by the end of 2020.
Installation will begin in 2021, followed by commissioning, with data
beginning in late 2023.



\acknowledgments

We are grateful for the vital contributions of the Fermilab staff
and the technical staff of the participating institutions.
This work was supported by the US Department of Energy; 
the Istituto Nazionale di Fisica Nucleare, Italy;
the Science and Technology Facilities Council, UK;
the Ministry of Education and Science, Russian Federation;
the National Science Foundation, USA; 
the Thousand Talents Plan, China;
the Helmholtz Association, Germany;
and the EU Horizon 2020 Research and Innovation Program under the
Marie Sklodowska-Curie Grant Agreement No.~690835 and 734303. 
This document was prepared by members of the Mu2e Collaboration
using the resources of the Fermi National Accelerator Laboratory
(Fermilab), a U.S. Department of Energy, Office of Science, HEP
User Facility. Fermilab is managed by Fermi Research Alliance, LLC
(FRA), acting under Contract No. DE-AC02-07CH11359.



\begin{thebibliography}{99}

\bibitem{TDR}
  L.~Bartoszek {\it et al.}, ``Mu2e Technical Design Report'', 
  arXiv:1501.05241, Fermilab-TM-2594, Fermilab-Design-2014-01.

\bibitem{Sindrum-II}
  W.H.~Bertl {\it et al.}, SINDRUM II Collaboration, ``A search
  for $\mu$--$e$ conversion in muonic gold'', Eur.~Phys.~J.~C 47
  (2006) 337.
  
\bibitem{CLFV-theory}
  A. de Gouv$\hat{\rm e}$a and P. Vogel, ``Lepton flavor and number
  conservation, and physics beyond the standard model'', Progress in
  Particle and Nuclear Physics 71 (2013) 75.

\bibitem{MELC}
  R.~Dzhilkibaev and V.~Lobashev, ``On the Search for $\mu\to e$
  Conversion on Nuclei'', Sov.~J.~Nucl.~Phys.~49 (1989) 384
  
\bibitem{Tracker}
  M.J.~Lee on behalf of the Mu2e collaboration, 
  ``The Straw-tube Tracker for the Mu2e Experiment'',
  Nuclear and Particle Physics Proceedings 273-275 (2016) 2530.

\bibitem{TDR-EMC}
  N.~Atanov {\it et al.}, ``The Mu2e Calorimeter Final Technical Design
  Report'', arXiv:1802.06341 (2018).

\bibitem{NIM-LYSO1}
  N.~Atanov {\it et al.}, ``Energy and time resolution of a LYSO matrix
  prototype for the Mu2e experiment'', Nucl.\ Instrum.\ Meth.\ A 824
  (2016) 684.

\bibitem{NIM-LYSO2}
  N.~Atanov {\it et al.}, ``Measurement of time resolution of the Mu2e
  LYSO calorimeter prototype'', Nucl.\ Instrum.\ Meth.\ A 812 (2016) 104.

\bibitem{NIM-BaF2}
  N.Atanov {\it et al.}, ``Design and status of the Mu2e electromagnetic
  calorimeter'', Nucl.~Instr.~Meth.~A 824 (2016) 695.
  
\bibitem{Proto-EMC}
  O.~Atanova {\it et al.}, ``Measurement of the energy and time
  resolution of a undoped CsI + MPPC array for the Mu2e experiment'', 
  JINST 12 (2017) P05007, arXiv:1702.03720.

\bibitem{Module-0}
  N.~Atanov {\it et al.}, ``Design and status of the Mu2e crystal
  calorimeter'', IEEE-TNS 65 (2018) 2073, arXiv:1802.06346.

\bibitem{BTF}
  G.~Mazzitelli {\it et al.}, ``Commissioning of the DA$\Phi$NE beam
  test facility '', Nucl.~Instr.~Meth.~A 515 (2003) 524.

\bibitem{QAcrystals}
  N.~Atanov {\it et al.}, ``Quality Assurance on Undoped CsI Crystals
  for the Mu2e Experiment”, IEEE 65 (2018) 752.

\bibitem{CRV}
  A.~Artikov {\it et al.}, ``Performance of Scintillator Counters
  with Silicon Photomultiplier Readout'', arXiv:1511.00374 (2015).
  
\bibitem{CRVnpe}
  A.~Artikov {\it et al.}, ``Photoelectron Yields of Scintillation
  Counters with Embedded Wavelength-Shifting Fibers Read Out With
  Silicon Photomultipliers'', Nucl.~Instr.~Meth.~A 890 (2018) 84,
  arXiv:1709.06587.

\bibitem{CRVradhard}
  G.~Blazey {\it et al.}, ``Radiation Tests of Hamamatsu Multi-Pixel
  Photon Counters'', Nucl.~Instr.~Meth.~A. 927 (2019) 463.
  
\end{thebibliography}
\end{document}